\documentclass{elsart}
\usepackage{graphicx}
\usepackage{dcolumn}
\usepackage{bm}
\usepackage{amsmath}
\usepackage{amsfonts}
\usepackage{amssymb}
\begin{document}
\begin{frontmatter}
\title{Quantum search algorithm as an open system }
\author{A. Romanelli, }
\author{A. Auyuanet\thanksref{PEPE}},
\author{and R. Donangelo\thanksref{UFRJ}}
\address{Instituto de F\'{\i}sica, Facultad de Ingenier\'{\i}a\\
Universidad de la Rep\'ublica\\ C.C. 30, C.P. 11000, Montevideo, Uruguay}
\thanks[PEPE]{Corresponding author. \textit{E-mail address:}
auyuanet@fing.edu.uy}
\thanks[UFRJ]{Permanent address: Instituto de F\'{\i}sica,
Universidade Federal do Rio de Janeiro\\
C.P. 68528, 21941-972 Rio de Janeiro, Brazil}
\date{\today}
\begin{abstract}
\vspace{0.2cm} \\
We analyze the responses of a quantum search algorithm to an external monochromatic 
field and to the decoherences introduced through measurement processes.
The external field in general affects the functioning of the search algorithm. 
However, depending on the values of the field parameters, there are zones where 
the algorithm continues to work with good efficiency. 
The effect of repeated measurements can be treated analytically and, in general, 
does not lead to drastic changes in the efficiency of the search algorithm.
\end{abstract}
\end{frontmatter}

\section{Introduction}

In the real world the concept of ``isolated system'' is an abstraction and
idealization. It was constructed to help understand some phenomena displayed
by real systems which may be regarded as approximately isolated. Since
dissipation is a macroscopic concept, there has been little interest in it
during the initial development of Quantum Mechanics. But, since about 40 years
dissipation has been incorporated into the quantum description to make
possible the understanding of processes such as ionization of atoms, radiation
fields inside a cavity or simply the decoherence caused by the interaction
between a system and its surroundings. The recent advances in technology that
have made possible to construct and preserve quantum states, have also opened
the possibility of building quantum computing devices
\cite{Dur,Sanders,Du,Berman}. Therefore, the study of the dynamics of open
quantum systems becomes relevant both for development of these technologies as
well as for the algorithms that will run on those future quantum computers.

Grover's quantum search algorithm \cite{Chuang}, which locates a marked item
in an unsorted list of $N$ elements in a number of steps proportional to
$\sqrt{N}$, instead of the $O(N)$ of the classical counterpart, is one of the
more studied. This search algorithm has also a continuous time version
\cite{Farhi} that has been described as the analog analogue of the original
Grover algorithm.

We have recently developed a new way to generate a continuous time quantum
search algorithm \cite{alejo}. In that work we have built a search algorithm
with continuous time that finds a discrete eigenstate of a given Hamiltonian
$H_{0}$, if its eigenenergy is given. This resonant algorithm behaves like
Grovers's, and its efficiency depends on the spectral density of the
Hamiltonian $H_{0}$. A connection between the continuous and discrete time
versions of the search algorithm was also established, and it was explicitly
shown that such a quantum search algorithm is essentially a resonance between
the initial and the searched state.

Recently \cite{Azuma,Saraceno,Shapira} the response of Grover's algorithm to
decoherences has been analyzed. Here a similar study on our proposal for a 
quantum search algorithm is performed. We rapidly review this method
in the following section, and then proceed to study how this resonant
algorithm can be affected by some interactions with the environment, as follows. 
In section 3, we subject the system to a monochromatic external field, and, in 
section 4, decoherences are introduced by performing a series of measurements. 
Conclusions are drawn in the last section of this work.

\section{Resonant Algorithm}

\label{sec:resonance} Let us consider the normalized eigenstates $\left\{
|n\rangle\right\}  $ and eigenvalues $\left\{  \varepsilon_{n}\right\}  $ for
a Hamiltonian $H_{0}$. Consider a subset \textbf{N} of $\left\{
|n\rangle\right\}  $ formed by $N$ states. Let us call $|s\rangle$ the unknown
searched state in \textbf{N} whose energy $\varepsilon_{s}$ is given. We
assume it is the only state in \textbf{N} with that value of the energy, so
knowing $\varepsilon_{s}$ is equivalent to ``marking'' the searched state in
Grover's algorithm.

In the resonant quantum search algorithm a potential $V$, that produces the
coupling between the initial and the searched states, is defined as
\cite{alejo}
\begin{equation}
V(t)=\left|  p\right\rangle \left\langle j\right|  \exp\left(  i\omega
_{sj}t\right)  +\left|  j\right\rangle \left\langle p\right|  \exp\left(
-i\omega_{sj}t\right)  \,, \label{potential}%
\end{equation}
where the eigenstate $|j\rangle$, with eigenvalue $\varepsilon_{j}$, is the
initial state of the system. This initial state is chosen so as not to belong
to the subset \textbf{N}. Above, $\left|  p\right\rangle \equiv\frac{1}%
{\sqrt{N}}{\displaystyle\sum\limits_{n\in{\mathbf{N}}}}|n\rangle$ is an
unitary vector which can be interpreted as the average of the set of vectors
in \textbf{N}, and $\omega_{sj}\equiv\varepsilon_{j}-\varepsilon_{s}$.

The objective of the algorithm is to find the eigenvector $|s\rangle$ whose
transition energy from the initial state $|j\rangle$ is the Bohr frequency
$\omega_{sj}$. In order to perform this task, the Schr\"{o}dinger's equation,
with the Hamiltonian $H=H_{0}+V(t)$, is solved. The wavefunction,
$|\Psi(t)\rangle$, is expressed as an expansion in the eigenstates
$\{|n\rangle\}$ of $H_{0}$, $|\Psi(t)\rangle=\sum_{m}a_{m}(t)\exp\left(
-i\varepsilon_{m}t\right)  |m\rangle$. The time dependent coefficients
$a_{m}(t)$ have initial conditions $a_{j}(0)=1$, $a_{m}(0)=0$ for all $m\neq
j$. We take units such that $\hbar=1$.

After solving this equation, the probability distribution results in,
\begin{align}
P_{j}  &  \simeq\cos^{2}(\Omega\ t)\ ,\nonumber\\
P_{s}  &  \simeq\sin^{2}(\Omega\ t)\ ,\label{psearched}\\
P_{n}  &  \simeq0,\text{ }n\neq j\text{ and }n\neq s,\nonumber
\end{align}
where $\Omega=\frac{1}{\sqrt{N}}$. From these equations it is clear that for
$\tau=\frac{\pi}{2\Omega}$ a measurement has a probability of yielding the
searched state very close to one. This approach is valid as long as all the
Bohr frequencies satisfy $\omega_{nm}\gg\Omega$ and, in this case, our method
behaves qualitatively like Grover's.

\section{Interaction with an external field}

The previous results have shown that the resonant search algorithm can be
reduced to essentially a quantum system with two energy levels. This reminds
us of the quantum optics problem of a two-level atom driven by a radiation
field, and the question arises on how the algorithm behaves when an external
field is interacting with it. Let us then consider the time-dependent
Schr\"{o}dinger equation
\begin{equation}
i\frac{\partial|\Psi(t)\rangle}{\partial t}=\left[  H_{0}+V(t)+\Gamma
(t)\right]  |\Psi(t)\rangle\,, \label{Schrodinger}%
\end{equation}
where the external monochromatic driving potential field%
\begin{equation}
\Gamma(t)=\Gamma_{0}\sin\omega_{0}t, \label{potential2}%
\end{equation}
where $\Gamma_{0}$ and $\omega_{0}$ are the field's amplitude and frequency,
respectively. Replacing the expansion for $|\Psi(t)\rangle$ in the equation
above, we obtain the set of equations for the amplitudes $a_{m}(t)$%
\begin{equation}
\frac{da_{n}(t)}{dt}=-i\sum_{m}\left\langle n|V(t)+\Gamma(t)|m\right\rangle
a_{m}(t)\exp\left(  -i\omega_{nm}t\right)  \,, \label{dinamics}%
\end{equation}
where $\omega_{nm}=\varepsilon_{m}-\varepsilon_{n}$ are Bohr frequencies.
Inserting the definitions (\ref{potential}) and (\ref{potential2}) into
(\ref{dinamics}), we find,
\begin{equation}
\frac{da_{n}(t)}{dt}=-i\sin\omega_{0}t\ \sum_{m}\Gamma_{nm}a_{m}(t)\exp\left(
-i\omega_{nm}t\right)  \,, \label{dinamics1}%
\end{equation}
for $n\notin\mathbf{N}$ and $n\neq j$, and%
\begin{align}
\frac{da_{n}(t)}{dt}  &  =-\frac{i}{\sqrt{N}}%
%TCIMACRO{\QATOPD{\{}{.}{{}}{{}}}%
%BeginExpansion
\genfrac{\{}{.}{0pt}{}{{}}{{}}%
%EndExpansion
\left(  1-\delta_{nj}\right)  a_{j}(t)\exp\left[  i\left(  \omega_{jn}%
+\omega_{sj}\right)  t\right] \nonumber\\
&  \makebox[2.5cm]{}\left.  +\delta_{nj}\sum_{m\in\mathbf{N}}a_{m}%
(t)\exp\left[  -i\left(  \omega_{jm}+\omega_{sj}\right)  t\right]  \right\}
\label{dinamics2}\\
&  -i\sin\omega_{0}t\ \sum_{m}\Gamma_{nm}a_{m}(t)\exp\left(  -i\omega
_{nm}t\right)  \ ,\nonumber
\end{align}
if $n\in\mathbf{N}$ or $n=j$. Above $\Gamma_{nm}=\left\langle n|\Gamma
_{0}|m\right\rangle $.

As in the general time-dependent perturbation problem, we are in the
presence of several time scales involved in the process. One is the fast 
scale associated to the Bohr frequencies, $\omega_{nm}$, another is 
the slow scale connected to the frequency $\Omega=1/\sqrt{N}$, related
to the natural oscilations of the system in the absence of an external field. 
There is also a third time scale, associated to the frequency $\omega_{0}$\ 
of the external field $\Gamma(t)$, which in principle could take any value.
We consider here cases where $\omega_{0}\lesssim\Omega$. In this case the 
presence of $\Gamma(t)$ is relevant, so we do not make a restriction on the 
range of validity of the algorithm. 

Following the general procedure of time-dependent perturbation theory, we 
integrate the previous equations over a time interval much greater than the 
one associated to the fast scale. In this way, terms having small phases 
dominate and the others average to zero. Therefore the equations become
\begin{align}
\frac{da_{j}(t)}{dt}  &  \simeq-\frac{i}{\sqrt{N}}a_{s}(t)-i\Gamma_{jj}%
a_{j}(t)\sin\omega_{0}t\ \ ,\nonumber\\
\frac{da_{s}(t)}{dt}  &  \simeq-\frac{i}{\sqrt{N}}a_{j}(t)-i\Gamma_{ss}%
a_{s}(t)\sin\omega_{0}t\ ,\label{dinamics3}\\
\frac{da_{n}(t)}{dt}  &  \simeq-i\Gamma_{nn}a_{n}(t)\sin\omega_{0}t\text{ for
all }n\neq s,j\ .\nonumber
\end{align}
These equations represent a pair of coupled oscillators, corresponding to the
initial and the searched state, subject to a harmonic external field, with the
initial conditions $a_{j}(0)=1$, $a_{n}(0)=0$ $(n\neq j)$. In such a case the
last equation above leads to $a_{n}(t)=0$ for all $n\neq s,j$. The remaining
two equations connect states $s$ and $j$. Making the change of variables%

\begin{align}
a_{j}(t) &  =x_{j}(t)\exp\left[  i\Gamma_{jj}\left(  \cos\omega_{0}t-1\right)
/\omega_{0}\right]  ,\nonumber\\
a_{s}(t) &  =x_{s}(t)\exp\left[  i\Gamma_{ss}\left(  \cos\omega_{0}t-1\right)
/\omega_{0}\right]  ,\label{cambio}%
\end{align}
in those equations, we obtain
\begin{align}
\frac{dx_{j}(t)}{dt} &  =-i\text{ }\Omega\text{ }x_{s}(t)\exp\left[
i\varepsilon\left(  \cos\omega_{0}t-1\right)  /\omega_{0}\right]  ,\nonumber\\
\frac{dx_{s}(t)}{dt} &  =-i\text{ }\Omega\text{ }x_{j}(t)\exp\left[
-i\varepsilon\left(  \cos\omega_{0}t-1\right)  /\omega_{0}\right]
\text{.}\label{newequa}%
\end{align}
where $\varepsilon=\Gamma_{ss}-\Gamma_{jj}$. Uncoupling these equations we
arrive at
\begin{align}
\frac{d^{2}x_{j}(t)}{dt^{2}}+i\varepsilon\sin\omega_{0}t\text{ }\frac
{dx_{j}(t)}{dt}+\Omega^{2}\text{ }x_{j}(t) &  =0,\nonumber\\
\frac{d^{2}x_{s}(t)}{dt^{2}}-i\varepsilon\sin\omega_{0}t\text{ }\frac
{dx_{s}(t)}{dt}+\Omega^{2}\text{ }x_{s}(t) &  =0.\label{endequ2}%
\end{align}
In the exceptional case when $\varepsilon=0$ the external monochromatic 
field does not affect the functioning of the search algorithm.
But, as a simple inspection of the differential equations (\ref{endequ2})
shows, when $\Gamma_{ss}\neq$\textbf{ }$\Gamma_{jj}$ and the ratio 
$\alpha=\omega_{0}/\Omega$ is irrational their solutions are not periodic.
This means that in general the application of the monochromatic field shall 
affect the functioning of the search algorithm. To be more quantitative, 
the last two equations, which differ only in the sign of the second term,
may be transformed into Hill's - type differential equations after a change 
of variables $x=\exp\left( i\varepsilon\int \sin\omega_{0}t\text{ }dt\right) y$, 
where $x$ is either $x_{j}(t)$ or $x_{s}(t)$. 
The solution of the resulting Hill's equations \cite{Hill} is a linear 
combination of the functions
\begin{align}
y_{1}(t) &  =\exp(\mu t)T_{1}(t),\label{Hilltion}\\
y_{2}(t) &  =\exp(-\mu t)T_{2}(t),\nonumber
\end{align}
where $T_{1}(t),\, T_{2}(t)$ are periodic functions with period $\pi$, 
and the Floquet exponent $\mu$ obeys the relation $\cosh\left( \pi\mu\right) =
\frac{1}{2}\left[ y_{1}(\pi)+\frac{dy_{2}}{dt}(\pi)\right]$. 
The initial conditions for $y_{1}(t)$ and $y_{2}(t)$ are $y_{1}(0)=1,\, y_{2}(0)=0;
\, \frac{dy_{1}}{dt}(0)=0,\, \frac{dy_{2}}{dt}(0)=1$. 
The stability of the solutions (\ref{Hilltion}) is determined by the values of 
the Floquet exponent $\mu$, which depends on the external field parameters. 
In particular, for zero or imaginary values of $\mu$, the solutions are stable. 

\begin{figure}[t]
\begin{center}
\includegraphics[scale=0.7]{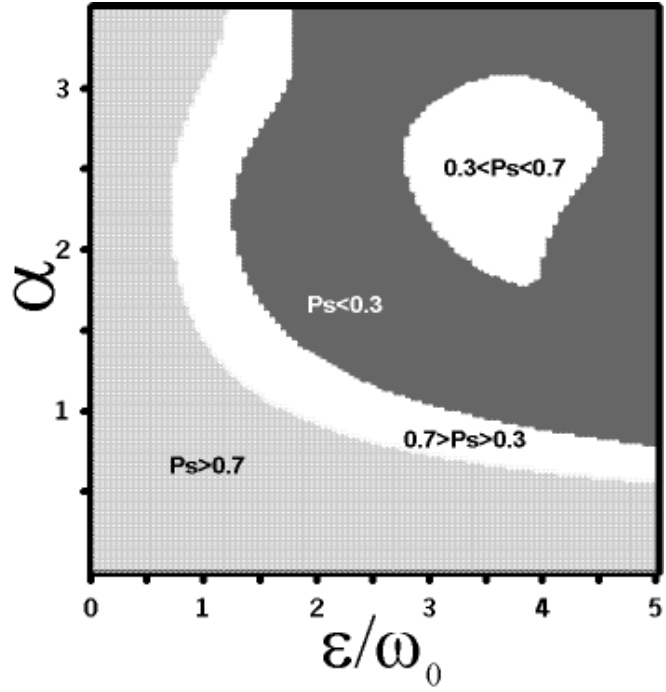}
\end{center}
\caption{{\footnotesize Probability of finding the searched state at the
optimal time as function of the field parameters $\alpha,\, \varepsilon$.}}
\label{fig:chance}%
\end{figure}

From these solutions one may obtain the probability of the searched state 
$P_{s}=\left|  a_{s}(t)\right| ^{2}$. Fig.~\ref{fig:chance} maps the value of this
probability, at the optimal time $\tau$, as a function of the parameters $\alpha$ 
and $\varepsilon$.
From it we may conclude that the search algorithm is useful also in the presence
of external field, since for an extense region of the field parameters it is possible 
to reach the searched state with high probability.

\section{Repeated measurements}

The search algorithm we are studying makes a transition from the initial to
essentially the sought state, other states being negligibly populated. Thus,
any measurement process will leave the system, with high probability, in one
of these two states. The probability distributions associated to the searched
state $|s\rangle$, and the initial state $|j\rangle$, initially evolve
according to the map of eq.(\ref{psearched}). If at time $t_{1}$ the state is
measured, the probabilities that the wavefunction collapses into either
$|s\rangle$ or $|j\rangle$ are given by
\begin{align}
P_{s}(t_{1})  &  \simeq\sin^{2}(\Omega\ t_{1})\ ,\nonumber\\
P_{j}(t_{1})  &  \simeq\cos^{2}(\Omega\ t_{1})\ , \label{prob}%
\end{align}
respectively. If after this first measurement the system is in state
$|j\rangle$ ($|s\rangle$), the probability distributions of the states
$|s\rangle$ and $|j\rangle$ after a second measurement, at a time $\Delta
t_{1}$ after $t_{1}$, are the eqs.(\ref{prob}), (the eqs.(\ref{prob})
exchanging $s$ and $j$), calculated at $\Delta t_{1}$. Between consecutive
measurements, the system undergoes an unitary evolution. Therefore, for
arbitrary intervals $\Delta t_{i}$ between consecutive measurements, the
probability distributions of the states $|s\rangle$ and $|j\rangle$ at a time
$t_{i+1}=t_{i}+\Delta t_{i}$ satisfy, always within the approximation
$\omega_{nm}\gg\Omega$, the matrix equations%
\begin{equation}
\binom{P_{s}(t_{i+1})}{P_{j}(t_{i+1})}=\left[
\begin{array}
[c]{cc}%
p_{i} & q_{i}\\
q_{i} & p_{i}%
\end{array}
\right]  \binom{P_{s}(t_{i})}{P_{j}(t_{i})}, \label{matrix}%
\end{equation}
where $p_{i}=\cos^{2}(\Omega\Delta t_{i})$ and $q_{i}=1-p_{i}$. This equation
is not a Markovian process because the transition probabilities are
time-interval dependent.

The general solution of the previous matrix equation for any time 
sequence of measurements is%
\begin{equation}
\binom{P_{s}(t_{m})}{P_{j}(t_{m})}=\left[
\begin{array}
[c]{cc}%
\alpha_{m} & \beta_{m}\\
\beta_{m} & \alpha_{m}%
\end{array}
\right]  \binom{P_{s}(0)}{P_{j}(0)}, \label{solution}%
\end{equation}
where $P_{s}(0)=0,P_{j}(0)=1$ and%
\begin{align}
\alpha_{m}  &  =\frac{1}{2}\left\{  1+\prod\limits_{i=0}^{m}\left[
2p_{i}-1\right]  \right\} \nonumber\\
\beta_{m}  &  =\frac{1}{2}\left\{  1-\prod\limits_{i=0}^{m}\left[
2p_{i}-1\right]  \right\}  \ . \label{coefficients}%
\end{align}

If we now consider that the measurement processes are performed at regular
time intervals, $t_{n}=n\Delta t$, eq.~(\ref{coefficients}) becomes
\begin{align}
\alpha_{m} &  =\frac{1}{2}\left[  1+\left(  \cos(2\Omega\Delta t)\right)
^{m}\right]  \nonumber\\
\beta_{m} &  =\frac{1}{2}\left[  1-\left(  \cos(2\Omega\Delta t)\right)
^{m}\right]  \ .\label{coefficients1}%
\end{align}
Now, we have an unitary evolution between consecutive measurements, and the
global evolution is a Markovian process, as all the $\Delta t_{i}$ are equal.

Let us now consider applications of the above eqs.(\ref{solution}%
-\ref{coefficients1}\textbf{).} In the case where the value of the optimal
time \textbf{$\tau$ }is only approximately known, we take a single measurement
at a time equal to our estimate of $\mathbf{\tau}$, which we designate as 
$\mathbf{\tau}^{\mathbf{\ast}}$. 
In that case the probability to find the searched state after the measurement 
is simply given by%
\begin{equation}
P_{s}(\mathbf{\tau}^{\mathbf{\ast}})=\frac{1}{2}\left[  1-\cos(2\Omega
\mathbf{\tau}^{\mathbf{\ast}})\right]  \simeq1-\frac{\pi^{2}}{4}\left(
\frac{\mathbf{\tau}^{\mathbf{\ast}}-\mathbf{\tau}}{\mathbf{\tau}}\right)
^{2}, \label{aprox}%
\end{equation}
which shows that the probability of finding the searched state, $P_{s}%
(\mathbf{\tau}^{\mathbf{\ast}}),$ decreases quadratically with the relative
error in the estimate of the optimal time.

\begin{figure}[t]
\begin{center}
\includegraphics[scale=0.45]{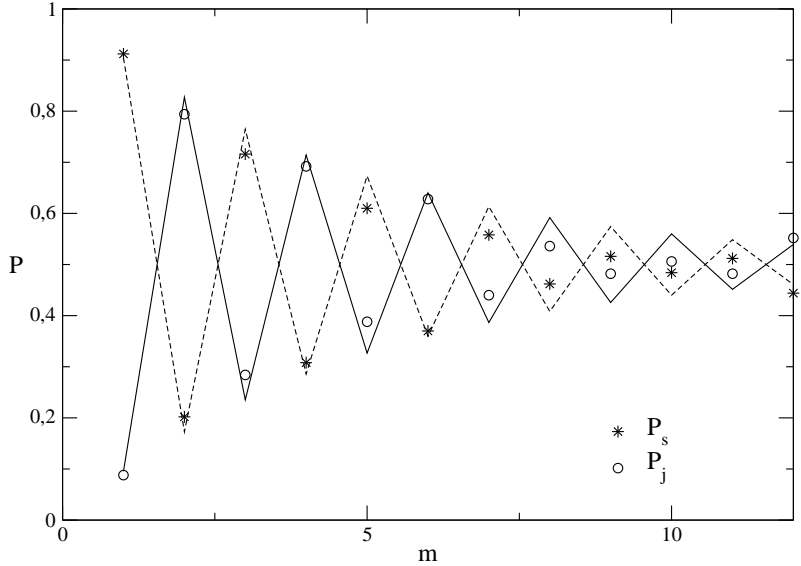}
\end{center}
\caption{{\footnotesize Probabilities of the searched ($P_{s}$) and the
initial ($P_{j}$) states as a function of the number of measurements ($m$),
for fixed $\Delta t=\tau^{\ast}$. The dashed and the full lines correspond 
to the theoretical development, eq.~(\ref{coefficients1}) in the same two
cases, respectively.}}%
\label{fig:sacando}%
\end{figure}

In Fig.~\ref{fig:sacando} we present the probabilities of the
searched and the initial states as a function of the number of measurements
$m$, in a case where $\Delta t=\tau^{\ast}$and the relative error
is$\ (\mathbf{\tau}^{\mathbf{\ast}}-\tau)/\tau=0.2$. The full and the dashed
lines correspond to the previous treatment, the circles ($P_{j})$ and the
stars ($P_{s}$) are the results obtained from a direct solution of the
Schr\"{o}dinger equation for the quantum rotor $H_{0}$, and for an ensemble of
$500$ trajectories with $N=50.$ We notice that after the first measurement the
value of $P_{s}$ is close to $0.9$. One also notices that after many measurements
this value tends to $0.5$.

We remark that eq.(\ref{coefficients1}) implies that for a
sufficiently large number of measurements, both $P_{s}$ and $P_{j}$ tend to
$1/2$, independently of the interval between measurements $\Delta t$ and the
initial conditions. Further inspection of Fig.~\ref{fig:sacando} shows that
the average value of $P_{s}$\ over different number of measurements $m$ is
also $1/2$.\ This means that the algorithm can be still useful, as it
may yield the searched state with a probability $0.5,$ even in the complete
absence of knowledge on the number of states on which the search is performed.
However, to perform this task in a reasonably small number of
measurements\ $m$, one has to make sure $\Delta t$\ is not too small, see
eq.(\ref{coefficients1}).

In the case where the $m$ measurements are performed in a total time
$\tau=\frac{\pi}{2\Omega}$, where the unperturbed algorithm converges to the
searched state, $\Delta t=\tau/m$, so the coefficients $\alpha_{m},\beta_{m}$
become
\begin{align}
\alpha_{m}  &  =\frac{1}{2}\left[  1+\left(  \cos\frac{\pi}{m}\right)
^{m}\right] \nonumber\\
\beta_{m}  &  =\frac{1}{2}\left[  1-\left(  \cos\frac{\pi}{m}\right)
^{m}\right]  \ . \label{coefficients2}%
\end{align}

\begin{figure}[t]
\begin{center}
\includegraphics[scale=0.45]{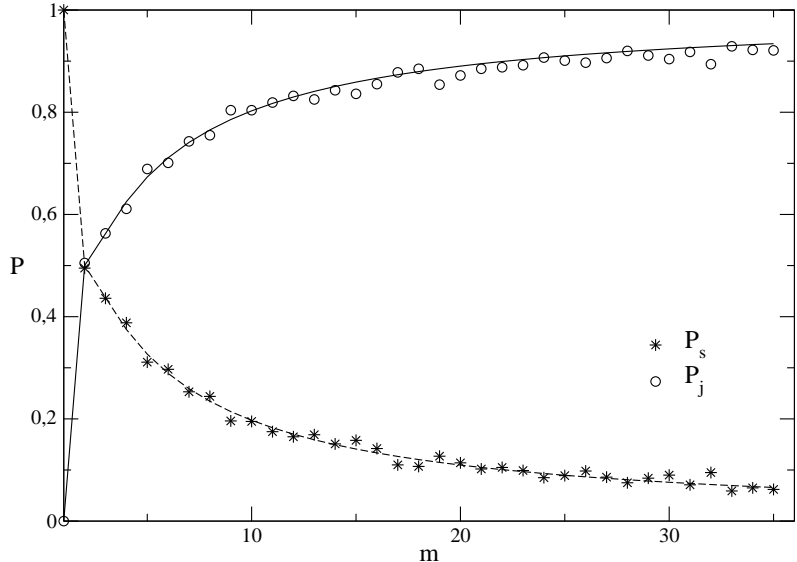}
\end{center}
\caption{{\footnotesize Probabilities of the searched ($P_{s}$) and the
initial ($P_{j}$) states as a function of the number of measurements ($m$),
performed in a total time $\tau$. The dashed and full lines are the theoretical 
results, obtained from eq.(\ref{coefficients2}), in both cases, respectively. }}%
\label{fig:midogro}%
\end{figure}

We show in Fig.~\ref{fig:midogro} the probabilities $P_{s}$ (stars), $P_{j}$
(circles) after $m$ measurements for $N=50$ and $500$ trajectories as in the
previous figure. The calculation based on the direct solution of the
Schr\"{o}dinger equation is compared to the one obtained from
eq.(\ref{coefficients2}), as a function of $m$. Besides the good agreement between
the two calculations we notice that $P_{s}$ decreases with $m,$ so that for
$m>30$, $P_{s}<0.1$.\ This simply means that the more frequently the wave
function collapses, the harder it becomes for the algorithm to significantly
depart from the initial state. Therefore, in this case the algorithm behaves
as an example of the Quantum Zeno effect, where the a high frequency of
measurements hinders the departure of the system from its initial state
\cite{Misra,Chiu}. This also explains our previous assertion, that for the
algorithm to be useful the time $\Delta t$ between measurements must not be
too small.

\section{Conclusions}

\label{sec:conclusion}

We have extended the study of the search algorithm presented in\ \cite{alejo}. 
There it was shown that the algorithm was robust when the energy of
the searched state had some imprecision. In this work, the resonant algorithm
was treated as an open system, and subject to two types of external
interactions: an oscillating external field and measurement processes.

It was shown that, although the algorithm is in general affected by a periodic
external field, for extensive zones of the field parameter values it works with 
good efficiency.
In the case of measurements, the probability distribution for the searched and 
initial states were obtained analytically. When a set of periodic measurements 
are performed, the probability distribution satisfies a Master equation. 
In this way we have shown that the global behavior of the algorithm becomes a 
Markovian process. Therefore, the algorithm can still be useful in the case of 
repeated measurements, as long as they are not too frequent.

We believe that these results may be directly extended to the Grover algorithm. 
As it may be seen from the description of this algorithm in ref.\cite{Chuang}, 
and as already pointed out by Grover \cite{Grover2}, this algorithm is essentially 
a resonance between the searched and the average state. One should also be aware 
that, while in the case of the Grover algorithm care is needed to make sure that 
after each measurement the state of the system is reinitialized into the average 
state, this procedure is not required in the algorithm proposed in this work.

\bigskip

We acknowledge the comments made by V. Micenmacher and the support from
PEDECIBA and PDT S/C/OP/28/84. R.D. acknowledges partial financial support
from the Brazilian National Research Council (CNPq) and FAPERJ (Brazil). A.R
and R.D. acknowledge financial support from the \textit{Brazilian Millennium
Institute for Quantum Information}.

\end{document}